\documentclass[aip,letterpaper, sort&compress, reprint]{revtex4-1}
\usepackage[T1]{fontenc}
\usepackage[latin9]{inputenc}
\usepackage{textcomp}
\usepackage{amsmath}
\usepackage{amssymb}
\usepackage{graphicx}
\usepackage{esint}
\usepackage{epstopdf}

\makeatletter

\@ifundefined{showcaptionsetup}{}{%

\PassOptionsToPackage{caption=false}{subfig}}
\usepackage{subfig}
\makeatother

\begin{document}

\title{Possible surface plasmon polariton excitation under femtosecond laser irradiation of silicon}

\author{Thibault J.-Y. Derrien}
\email{thibault.derrien@gmail.com}
\affiliation{Laboratoire Hubert Curien (LabHC), UMR CNRS 5516 -
Universit\'{e} Jean-Monnet. \\
B\^{a}timent F, 18 rue du Professeur Benoit Lauras, F-42000 Saint-Etienne, France.}
\affiliation{Laboratoire Lasers, Plasmas et Proc\'{e}d\'{e}s Photoniques (LP3), UMR CNRS 7341 - Aix-Marseille Universit\'{e}, \\Parc Technologique et Scientifique de Luminy, Case 917, 163 avenue de Luminy, F-13288 Marseille CEDEX 09, France.}

\author{Tatiana E. Itina}
\affiliation{Laboratoire Hubert Curien (LabHC), UMR CNRS 5516 -
Universit\'{e} Jean-Monnet. \\
B\^{a}timent F, 18 rue du Professeur Benoit Lauras, F-42000 Saint-Etienne, France.}

\author{R\'{e}mi Torres}
\affiliation{Laboratoire Lasers, Plasmas et Proc\'{e}d\'{e}s Photoniques (LP3), UMR CNRS 7341 - Aix-Marseille Universit\'{e}, \\Parc Technologique et Scientifique de Luminy, Case 917, 163 avenue de Luminy, F-13288 Marseille CEDEX 09, France.}

\author{Thierry Sarnet}
\affiliation{Laboratoire Lasers, Plasmas et Proc\'{e}d\'{e}s Photoniques (LP3), UMR CNRS 7341 - Aix-Marseille Universit\'{e}, \\Parc Technologique et Scientifique de Luminy, Case 917, 163 avenue de Luminy, F-13288 Marseille CEDEX 09, France.}

\author{Marc Sentis}
\affiliation{Laboratoire Lasers, Plasmas et Proc\'{e}d\'{e}s Photoniques (LP3), UMR CNRS 7341 - Aix-Marseille Universit\'{e}, \\Parc Technologique et Scientifique de Luminy, Case 917, 163 avenue de Luminy, F-13288 Marseille CEDEX 09, France.}

\begin{abstract}
The mechanisms of ripple formation on silicon surface by femtosecond
laser pulses are investigated. We demonstrate the transient evolution
of the density of the excited free-carriers.
As a result, the experimental conditions required for the excitation
of surface plasmon polaritons are revealed. The periods of the resulting
structures are then investigated as a function of laser parameters,
such as the angle of incidence, laser fluence, and polarization. The
obtained dependencies provide a way of better control over the properties
of the periodic structures induced by femtosecond laser on the surface
of a semiconductor material.
\end{abstract}


\maketitle
\thispagestyle{empty}

\section{Introduction}

Femtosecond lasers are known to be powerful tools for micro- and nanomachining. In particular, these lasers can induce periodic modulations ("Laser-Induced Periodic Surface Structures" or LIPSS) on the surfaces of metals, semiconductors and dielectric samples at relatively moderate laser fluences \cite{Birnbaum1965,Young1983,Ursu1985,Bonse2010, Tsibidis2012,Renger2009}. Furthermore, it is possible to decrease periods of these structures below the laser wavelength, thus rising the precision of laser nanomachining beyond the diffraction limit \cite{Borowiec2003,Bonse2005,Vorobyev2008a,Crawford2007}.
Applications of the ripple structures are numerous. For instance,
it is possible to colorize metals and to control over laser marking \cite{Vorobyev2008,Vorobyev2008b}.

For further development of these applications, it is important to
better understand the mechanisms of ripple formation by femtosecond laser pulses.
The number of the observed structures is, however, very large making the control over the laser
parameters very complicated. In general, two types of structures can be distinguished (i) resonant structures,
where the resulting period is correlated with laser wavelength,
and (ii) non-resonant structures, which are not explicitly connected
with the laser wavelength and with the coherent effects. Thus, the
so-called Low-Spatially-Frequency LIPSS (LSFL) belong to the
resonant structures \cite{Bonse2005}. In addition, much larger parallel
ripples of several micrometers, can be considered to belong to the
non-resonant structures. In this case, a large number of pulses is
required to provide a thick melted depth comparable to the structure
amplitude, so that these ripples are rather connected with the capillary
wave generation or surface stress \cite{Tsibidis2012}. The non-resonant structures were also explained
by the self-organized processes, described by the \emph{Kuramoto-Sivashinsky
}equation \cite{Reif2002,Ben-Yakar2007,Varlamova2007}.
With the increase in laser pulses and fluence,  drop-like structures named ''beads'', and then conical structures called "black silicon" can be also obtained \cite{Her1998, Carey2003, Sarnet2008a,Bonse2010,Reif2010}.

In this paper, we focus our attention at the resonant periodic structures
with the period near the laser wavelength e.g. the LSFL. The classical
theory of ripple formation proposes that scattering of the laser wave
by surface roughness couples the laser wave with the surface modes,
which interfere with the laser light, and thus lead to a periodic modulation
of the absorbed energy \cite{Sipe1983}.
This theory was recently confirmed by the numerical calculations based
on the system of Maxwell equations for rough surface \cite{Skolski2012}.
This scenario requires the presence of an initial roughness of a certain
size. Both the laser parameters and the
surface roughness  are, however, often unknown. In addition, several laser pulses are frequently required
to form a structure. In this case, the period of the
energy deposition can be smaller than laser wavelength, as was
explained by Bonse et al \cite{Bonse2009} by using the ''Sipe-Drude'' model.
In this way, an explanation of the very narrow structures (HSFL) appearing
after numerous pulses \cite{Costache2004,Reif2006a} was proposed. 
In addition, Tsibidis et al. \cite{Tsibidis2011} recently investigated the cumulative hydrodynamic effects and the corresponding surface modifications. It was found that a non-resonant mechanism explains the reduction of the LSFL periodicity with the increase of pulse number in Si.
In addition, the possibility to create periodic
surface modifications with a single femtosecond pulse was demonstrated
for metals and semiconductor materials \cite{Bonse2009,Guillermin2007,Sarnet2008a}.
To explain the formation of the near-wavelength ripples at intense
and reduced number of pulses, several authors have proposed the surface
plasmon polaritons (SPP) as the mechanism responsible for the surface
wave generation in semiconductors and dielectrics \cite{Miyaji2008, Huang2009a, Bonse2009}. The surface
plasmon polaritons are known to be excited on metal surfaces.
The ripple formation mechanism for metals has been linked with the
excitation of surface plasmon polariton by several authors \cite{Sakabe2009, Garrelie2011}. However,
the excitation conditions remain rather puzzling in the case of semiconductor
or dielectric materials. Moreover, it is not clear if the SPP coupling
can occur by using a single femtosecond laser pulse, since specific
coupling conditions are required to add the missing momentum at the
surface. In particular, gratings, snom probe, prism, defects, or roughness
\cite{Raether1986,Zayats2005} typically help to couple
the laser wave with a surface wave mode. Thus, our study focuses on the semiconductor case, and the analyzed material is monocrystalline Si.

In the case of semiconductor materials, laser-induced modification
of the dielectric function changes material properties. As a result,
the importance of a transitory metallic state was underlined by Bonse et al \cite{Bonse2011}.
In such cases, however, the presence of a nanometric defect (such
as bubbles or nanoparticle) at the surface is required and those methods
correspond to the well-known case of localized surface plasmon (LSP)
excitation around isolated defects \cite{Hecht1996}.
It was demonstrated, furthermore,
that the transient modification of the solid properties follows the
plasma dynamics of the free-carrier gas, due to their excitation by
the intense laser \cite{Sokolowski-Tinten1998,Sokolowski-Tinten2000}.
However, a systematic study is still required for the conditions of
SPP excitation on semiconductor surface by a femtosecond laser interaction.
Under these excitation, electron-hole pairs are generated, but
also thermal effects play a role. That is why a clear explanation
is required for the laser parameter range, ambient environment and
sample surface conditions. To help developing the corresponding applications,
the resulting ripple periods should be connected with laser parameters.
In this paper, we consider the conditions required to excite surface
plasmon polaritons on semiconductor's surface. The developed model
provides all the parameters, which lead to the SPP excitation, such
as angle of incidence, laser fluence, pulse duration, and surface roughness
for the given ambient optical properties. It is demonstrated that,
under the required conditions, Si surface becomes optically active
under femtosecond irradiation, and thus, SPPs can be excited by irradiation
of a coupling device. The resulting periods are analyzed as a function
of laser parameters.

The paper is organized as follows. In Section \ref{sec:experiments}, we present the experimental
protocol. In Section \ref{sec:Modeling-of-the}, we present the model and consider the modification
of the optical properties of Si under femtosecond irradiation. In Section \ref{sec:Polaritons},
conditions of the excitations of the surface plasmon polaritons are presented. In Section
\ref{sec:Results}, the laser parameters allowing the SPP excitation are examined as a function of
laser fluence and laser pulse duration. Then, the required minimal roughness thickness is
discussed. Finally, the calculated periodicities are compared to the experimental values as a
function of laser fluence. The evolution of the ripple period is analyzed as a function of angle
of incidence and laser polarization.

\section{\label{sec:experiments}Experimental details}

The micromachining experiments were performed by using a Ti-sapphire laser
(Hurricane model, Spectra-Physics) that was operated at 800 nm, with an energy of 500 $\mu J$, a
repetition rate of 1 kHz and a laser pulse duration of 100 fs. Laser irradiation of silicon surface was carried out in a vacuum system with a pressure of $5 \times 10^{-5}$
to $1 \times 10^{-5}$ mbar. This low pressure considerably reduces the redeposition of unwanted debris from the
laser ablation process.  
To get a more uniform laser energy
distribution, only the center part of the gaussian laser beam was selected using a square mask of $2
\times 2$ mm$^{2}$. A spot about $35 \times 35$ $ \mu m^{2}$ area was obtained projecting the mask image onto the sample
surface with a lens (f' = 50 mm). Laser beam was perpendicular to the sample surface. In the present study, we consider only linear polarization. The laser
energy delivered to the sample surface could be attenuated by coupling an analyzer and a polarizer
and completed by a set of neutral density filters. The analyzer rotation placed in front of
the polarizer is controlled by a computer. The engraving results are in situ monitored by a CCD
camera. The number of pulses is controlled by triggering a Pockels cell, thus reducing the repetition rate of the laser pulse to 5
Hz. 
We irradiated a <100> monocrystalline silicon (c-Si) wafer by one or several laser
pulses at fluences of $0.5$ $J/cm^{2}$, $0.8$ $J/cm^{2}$, and $1.15$ $J/cm^{2}$.
Two series of experiments were performed. (i) At very low (one or two) number of pulses, the angle of incidence has been kept normal to the surface. (ii) At $N=10$ pulses, the angle of incidence and the laser polarization have been varied.

\section{\label{sec:Modeling-of-the}Modeling details}

Femtosecond laser can promote carriers from the valence band of a semiconductor to the conduction
band leading to free-carrier absorption. In our model, the number
density of the carriers in the conduction band is calculated by solving
the following equation
\begin{flalign}
\frac{\partial n_{e}}{\partial t}-\boldsymbol{\nabla}\cdot\left(k_{B}T_{e}\mu_{e}\boldsymbol{\nabla}n_{e}\right) & =G_{e}-R_{e}\label{eq:ElectronDensity}
\end{flalign}
where $n_e$ is free-carriers number density, and $G_{e}=\left[\frac{\sigma_{1}I}{\hbar\omega}+\frac{\sigma_{2}I^{2}}{2\hbar\omega}+\delta_{I}n_{e}\right]\frac{n_{v}}{n_{e}+n_{v}}$
is the gain of free-carriers per unit time and unit volume ($m^{-3}.s^{-1}$). $n_v$ is the quantity of valence band electrons.

\begin{table*}
\begin{centering}
\begin{tabular}{cc|cc|c}
{\small Physical meaning} & {\small Notation} & {\small Value} & {\small Unit} & {\small Reference}\tabularnewline
\hline
\hline
{\small One-photon absorption coefficient} & {\small $\sigma_{1}$} & {\small $2\omega Im\sqrt{\varepsilon_{\infty}\left(\omega\right)}/c=1.021 \times 10^{5}$} & {\small $\mathrm{m^{-1}}$} & {\small (d, g)}\tabularnewline
{\small Two-photon ionization rate} & {\small $\sigma_{2}$} & {\small $0.1 \times 10^{-9}$} & {\small $\mathrm{m.W^{-1}}$} & {\small (c)}\tabularnewline
{\small Impact ionization probability rate} & {\small $\delta_{I}$} & {\small $3.6 \times 10^{10}e^{-E_{g}/k_{B}T_{e}}$} & {\small $\mathrm{s^{-1}}$} & {\small (b, e, f)}\tabularnewline
{\small Auger recombination rate} & {\small $C$} & {\small $3.8 \times 10^{-43}$} & {\small $\mathrm{m^{6}.s^{-1}}$} & {\small (b)}\tabularnewline
{\small Recombination delay at high density} & {\small $\tau_{0}$} & {\small $6$} & {\small ps} & {\small (a)}\tabularnewline
\end{tabular}
\par\end{centering}

\caption{\label{tab:DensityOfFluence} The calculation parameters for c-Si
under 800 nm irradiation. References: (a) Ref. \cite{Bok1981}, (b)
Ref. \cite{Driel1987}, (c) Ref. \cite{Sjodin1998}, (d) Ref. \cite{Palik1998},
(e) Ref. \cite{Sze2007}, (f) Ref. \cite{Thornber1981}, (g) Ref.
\cite{Choi2002a}.}
\end{table*}
Both one-photon inteband cross-section ($\sigma_{1}$)
and the two-photon cross-section ($\sigma_{2}$) are used in the model (Table \ref{tab:DensityOfFluence}).
The conduction band can be also populated due to the electron impact ionization (avalanche process).
The corresponding coefficient $\delta_{I}$ is also given in Table \ref{tab:DensityOfFluence}.
$R_{e}=\frac{n_{e}}{\tau_{0}+\frac{1}{Cn_{e}^{2}}}$ is
the loss of conduction electrons by Auger recombination, where the recombination time $\tau_{\text{0}}$
is equal to 6 ps in our calculations\cite{Bok1981, Bulgakova2005}.

The initial density
of free-carriers present in the conduction band is
$1.84 \times 10^{9}\, cm^{-3}$ at a temperature of $300$ K. $k_{B}$ is the Boltzmann
constant. In the near-ablation regime, using low energy photons (1.5 eV in our case), the number of excitable electrons is limited to the ones available in the valence band. Even in ablation regime at the considered laser intensities, less than one electron per atom is usually promoted to the conduction band\cite{Bulgakova2005}. Thus, the number of the excitable valence band electrons is described by $n_{0}=\rho_{Si}=5 \times 10^{22}\,cm^{-3}$, equal to the density of the Si lattice. During the excitation, the number of valence band electrons $n_v$ is therefore calculated by $n_v=n_0-n_e$.

The free-carrier mobility is described by $\mu_{e}=\frac{e}{m_{e}\nu}$
where $\nu=1.5 \times 10^{14}\,s^{-1}$ is the free-carrier collision frequency. Collision frequency is adjusted in agreement with melting fluence and melted depth given by Bonse et al \cite{Bonse2004}, consistent with Monte Carlo simulations of collision frequency in Si\cite{Fischetti1988}. $m_{e}$ is the optical mass of electron-hole pairs, which is equal to \cite{Sokolowski-Tinten2000} $m_{e} = 0.18 m_{e0}$, where $m_{e0}$ is the electron mass.
The density of electrons is calculated
by using Eq. (\ref{eq:ElectronDensity}) taking into account thermal
diffusion and Auger recombination. The hole temperature and density are
considered equal to the ones of free-electrons, since the contribution
of the electron-hole pairs to the absorption is taken into account
by the optical mass in dielectric function.

Laser energy absorption is calculated
as follows \cite{Driel1987,Bulgakova2005}

\begin{equation}
\frac{\partial I}{\partial z}=-\alpha_{fcr}I-(\sigma_{1}I+\sigma_{2}I^{2})\frac{n_{v}}{n_{e}+n_v},\label{eq:Beer-Lambert-1}
\end{equation}
where $I$ is the local intensity. Intensity at the surface is given
by $I_{z=0}(t,x)=\left[1-R\left(x\right)\right]I_{0}(t,x)\frac{1}{\cos\theta}$
and $I_{0}(t,x)=\frac{2F}{\tau}\sqrt{\frac{\ln(2)}{\pi}}e^{-\frac{1}{2}\left(\frac{x}{\sigma_{x}}\right)^{2}}e^{-\frac{1}{2}\left(\frac{t-t_{0}}{\sigma_{\tau}}\right)^{2}}$.
$F$ denotes the maximum fluence reached during the interaction. $t_{0}=0$ in our calculations. Spot size $w_0$ and pulse duration $\tau$ are respectively defined
at the FWHM of spatial and temporal gaussian distributions. Thus,
$\sigma_{x}=\frac{w_{0}}{2\sqrt{2\ln2}}$ and $\sigma_{\tau}=\frac{\tau}{2\sqrt{2\ln2}}$. The free-carrier absorption is described by $\alpha_{fcr}=\frac{2\omega n_{2}}{c}$ where $n_{2}=\sqrt{0.5 \left(-\Re e\left(\varepsilon_{fcr}\right)+\sqrt{\Re e\left(\varepsilon_{fcr}\right)^{2}+\Im m\left(\varepsilon_{fcr}\right)^{2}}\right)}$ and $\varepsilon_{fcr} = 1 - \frac{\omega_{p}^{2}}{\omega^{2}} \frac{1}{1+i\frac{\nu}{\omega}}$ describes the dielectric response of the free-carriers.

During the interaction, the surface reflectivity changes depending
on the angle of incidence $\theta$ and on laser polarization. In the present study, we consider only linear laser polarization. 

We take into account the contribution
of each layer to the surface reflectivity. The simulated region is considered as a layered structure where each layer $j$ has its own optical properties related to the local free-carrier density. The layer thickness $h_j$ coincides with the numerical cell width. The fraction of the plain wave amplitude reflected by the layers from $j$ to $k$ is calculated by the recurrence formula 
\[
r_{j,k}=\frac{r_{j,j+1}+r_{j+1,k}e^{2i\phi_{j+1}}}{1+r_{j,j+1}r_{j+1,k}e^{2i\phi_{j+1}}}.
\]
The final value of the surface reflection coefficient $R$ is determined as $R=\left| r_{0,N} \right|^2$ where $0$ corresponds to the sample-vacuum interface (with optical properties of the environment) and $N$ is a layer located deep in the substrate. $\phi_{j+1}=\frac{2\pi h_{j+1}}{\lambda}\sqrt{\varepsilon_{j+1}}$
is the phase shift induced by the $j+1$ layer, $h_{j+1}$ is the
thickness of the layer $j+1$. 
$r_{j,j+1}$ is given by Fresnel equations. \cite{Jackson1999} $\varepsilon_{j}$ is the
full dielectric function given by Eq. (\ref{eq:fulldielectricfunction})
calculated at the layer $j$. The dielectric function
is calculated as follows \cite{Sokolowski-Tinten2000,Bulgakova2005}

\begin{equation}
\varepsilon_{Si}(\omega,n_{e},\nu)=1+\left(\varepsilon_{\infty}(\omega)-1\right)\frac{n_{v}}{n_{e}+n_v}-\frac{\omega_{p}^{2}}{\omega^{2}\left(1+i\frac{\nu}{\omega}\right)}\label{eq:fulldielectricfunction}
\end{equation}
where $\varepsilon_{\infty}\left(\omega\right)$ is the dielectric
constant and depends on laser wavelength \cite{Palik1998}. $\omega_{p}$ is the plasma pulsation defined here by $\omega_{p}^{2}=\frac{n_{e}e^{2}}{m_{e}\varepsilon_{0}}$.
$\omega$ is the laser pulsation defined by $\omega=\frac{2\pi c}{\lambda}$.
Photo-emission and thermo-emission are neglected, since Dumber
field strongly limits the transport and thus the loss of free-carriers
at the surface. \cite{Bulgakova2005, Bulgakova2010}
%

Several parameters of the above model depend on temperature. Under femtosecond laser irradiation, the
electron-hole sub-system is excited to much higher temperatures and electron heat conductivity is
much larger than that of lattice. To calculate temperatures of carriers and lattice, a
two-temperature model is used. The laser-excited zone is small, so that ballistic effects should be
also accounted for the electron sub-system. Therefore, we use the following ballistic-diffusive
equation \cite{Jou1993,Alvarez2007,Palpant2008} to describe the electron-hole subsystem.

\begin{align*}
\frac{1}{4\nu}\frac{\partial^{2}T_{e}}{\partial t^{2}}+\frac{\partial T_{e}}{\partial t} & =\boldsymbol{\nabla}\left(D_{SBD}\boldsymbol{\nabla}T_{e}\right)-\frac{\gamma_{ei}}{C_{e}}\left(T_{e}-T_{i}\right)+\frac{Q_{e}}{C_{e}},
\end{align*}
where $D_{SBD}=\frac{L^{2}\nu}{6\pi^{2}}\left[\sqrt{1+4\pi^{2}Kn^{2}}-1\right]$
is the effective free-carrier thermal diffusion term, based on the
ratio $Kn=\frac{l_{MFP}}{L}$ of free-carrier mean-free-path $l_{MFP}=\frac{1}{\nu}\sqrt{\frac{3k_{B}T_{e}}{m_{e}}}$
and size of the excited zone $L\sim\left(2\omega \Im m \sqrt{\varepsilon_{Si}}/c\right)^{-1}$.
Specific heat capacity of free-carriers is taken to be equal to the
classical limit $C_{e}=\frac{3}{2}k_{B}n_{e}$. The source term $Q_{e}
=\left[\left(\hbar\omega-E_{g}\right)\frac{\sigma_{1}I}{\hbar\omega}+\left(2\hbar\omega-E_{g}\right)\frac{\sigma_{2}I^{2}}{2\hbar\omega}-E_{g}\delta_{I}n_{e}\right]\frac{n_{0}-n_{e}}{n_{0}}
+\alpha_{fcr}I+E_{g}R_{e}-\frac{3}{2}k_{B}T_{e}\frac{\partial n_{e}}{\partial t}$
describes the energy of free-carriers by taking into account one-photon
and two-photon ionization, the energy loss by electron avalanche,
the free-carrier heating, the release of energy due to Auger recombination
and last term comes from the variation of the specific heat capacity
with time, since density is strongly modified during the pulse.

The time of free-carrier coupling to the lattice  was experimentally determined by Sjodin et al \cite{Sjodin1998} as a function of free-electron density given by $\tau_{\gamma}=\tau_{\gamma0}\left[ 1+\left(\frac{n_{e}}{n_{th}} \right)^{2} \right] $ with $n_{th}=6.02 \times 10^{20}\ cm^{-3}$
and $\tau_{\gamma0}=240$ fs. In our calculations, the coupling rate is given by \cite{Bulgakova2005} $\gamma_{ei}=\frac{C_{e}}{\tau_{\gamma}}$ .
Because of the slow thermal diffusion of the lattice energy, we describe the temperature of the lattice $T_{Si}$ by a classical diffusion equation, taking into account the energy transfered from free-carriers as follows

\[
C_{Si}\frac{\partial T_{Si}}{\partial t}=\boldsymbol{\nabla}\left(\kappa_{Si}\boldsymbol{\nabla}T_{Si}\right)+\gamma_{ei}\left(T_{e}-T_{Si}\right)
\]

The specific heat capacity of Si is a function of liquid density fraction. For solid state,
dependence with temperature is given by relations \cite{Driel1987} $C_{s-Si}[J.m^{-3}]
=10^{6}\left[1.978+3.54 \times 10^{-4}.T-3.68T^{-2}\right]$ and $\kappa_{s-Si}[W/m/K] =10^{2}\left[1585
T^{-1.23}\right]$. For liquid state, parameters are given by \cite{Desai1986,Rhim2000,Magna2004}
$C_{l-Si}(T) =1.045 \times 10^{3}\rho_{l-Si}$, where $\rho_{l-Si}=2520$ $kg/m^{3}$ and $\kappa_{l-Si}(T)
=10^{2}\left[0.502+29.3 \times 10^{-5}\left(T-T_{m}\right)\right]$. Melting temperature $T_{m}$ depends on
the free-electron density as described by Eq. (\ref{eq:MeltingTemperature}). During the phase
transition, both Si heat capacity and conductivity are calculated using the fraction of liquid
$\eta$, and are respectively defined by $C_{Si}(T)=\left(1-\eta\right)C_{s-Si}(T)+\eta C_{l-Si}(T)$
and $\kappa_{Si}(T)=\left(1-\eta\right)\kappa_{s-Si}(T)+\eta\kappa_{l-Si}(T)$. Melting is considered
by using melting enthalpy $\Delta H_{m}$ at the melting temperature $T_{m}$. The resulting thermal
energy is given by $U=\int_{T_0}^{T_m} C_{s-Si}(T')dT' + \Delta H_m + \int_{T_m}^{T}
C_{l-Si}(T')dT'$ where $U$ is the internal energy of the lattice, $T$ is the Si temperature,
$T_0=300$ K is the initial temperature of the system, and $\Delta H_m=4 \times 10^{9}$ $J/m^{3}$ is the
melting enthalpy of Si.

%
%

Previously, two different types of phase transitions were shown to take place for Si
\cite{Rousse2001,Sundaram2002,Sokolowski-Tinten2003}. The first one is  thermal melting
 and is described by a thermal criterion based on the required energy $E_{m}=k_{B}T_{m}\rho_{Si}+\Delta H_{m}$. 
The second phase transition mechanism is a so-called ''non-thermal melting''  due to
the lattice decomposition due to a large number of carriers in the conduction band.
The contribution of the non-thermal melting is also taken into account
by the decrease of the band gap energy as a function of free-carrier
density \cite{Driel1987} (limited to positive or null values) expressed by
$E_{g}(T,n_{e})=1.17-4.73 \times 10^{-4}\frac{T^{2}}{T+636}-1.5 \times 10^{-10}n_{e}^{1/3}$
and by a lowering of the melting temperature
described by the relation\cite{Combescot1985}

\begin{equation}
T_{m}=T_{m}^{0}-\frac{n_{e}E_{gap}}{C_{s-Si}.}\label{eq:MeltingTemperature}
\end{equation}
where $T_m^0=1687$ K. 

Boundary conditions for transport equations are set so that free-carriers do not leave the
sample. The sample is 250 \textmu{}m thick, and the optical transmission has been checked to be zero trough the sample.

\section{\label{sec:Polaritons}Surface Plasmon Polaritons}

Light can be coupled from free space into the surface plasmon polaritons (SPP) only by matching the momentum of the SPPs.
This can be done via index matching \cite{Raether1986}, or grating coupling \cite{Ursu1984,Bonch-Bruevich1992}.
In addition, other cases can be considered.  A
non-resonant excitation can be performed by scattering of the laser wave on surface defects or a
surface roughness. In such a case, laser wave is scattered on a broad angular distribution, and a part of the laser energy
couples with surface modes. Laser wave can also interact with near-wavelength structures as
described by Mie scattering, which leads to the excitation of localized surface plasmons (LSP) \cite{Hecht1996,Novotny1997}.

In each case, the excitation of surface waves requires several resonance
conditions \cite{Raether1986,Zayats2005,Maier2007}.
In this part, we present theoretical conditions allowing the excitation
of the Surface Plasmon Polaritons (SPPs) at the laser-irradiated surface of Si.

The excitation conditions of surface plasmon polaritons  at a flat surface is that
the corresponding curves cross in the dispersion diagram \cite{Raether1986,Maier2007}.
The dispersion relation is obtained from  the boundary
conditions of the electric and magnetic field at the interface. The
continuity of the electric field at the interface results to the expression
$\frac{k_{2}}{k_{1}}=-\frac{\varepsilon_{2}}{\varepsilon_{1}}$, where
$\varepsilon_{1,2}$ are the dielectric constants on both sides of
the interface, and $k_{1,2}$ are the respecting momenta of the both
sides of the interface. In the general case, this expression can be
verified only if
\begin{equation}
\Re e\left(\varepsilon_{1}\right)\Re e\left(\varepsilon_{2}\right)<0\label{eq:SPPflatCondition1}
\end{equation}
 which corresponds to a metal-dielectric interface. The continuity
of the magnetic field at the interface leads to the expression of
the SPP wave-number
\begin{equation}
\beta=\frac{\omega}{c}\sqrt{\frac{\varepsilon_{1}\varepsilon_{2}}{\varepsilon_{1}+\varepsilon_{2}}}
\label{eq:SPPwavenumber}
\end{equation}
 and to a second SPP excitation condition given by:
\begin{equation}
\Re e\left(\varepsilon_{2}\right)<-\Re e\left(\varepsilon_{1}\right) \label{eq:ConditionMagField}
\end{equation}

\begin{figure} \begin{centering} \includegraphics[width=9cm]{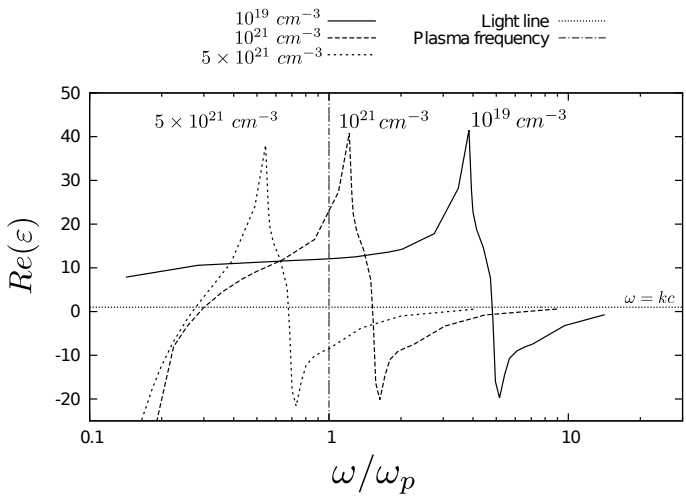}
\par\end{centering} \caption{\label{fig:DielectricFunctionWithNe}Real part of the dielectric
function of Si $\varepsilon (\omega)=\left( \frac{k c}{\omega} \right)^{2} $ as a function of photon energy compared to the plasma frequency
$\omega_{p}=\sqrt{\frac{n_{e} e^{2}}{m_{e} \varepsilon_{0}}}$. If the plasma frequency is located on the left of the resonance peak, surface mode is not bounded to the surface and SPP are not allowed. If the plasma frequency is located on the right of the surface plasmon polariton resonance peak, the SPPs are allowed.} \end{figure}

By introducing Eq. (\ref{eq:fulldielectricfunction}) into Eq. (\ref{eq:SPPwavenumber}) and
considering an interface with vacuum ($\varepsilon_{1}=1$), we calculate the dispersion relation as
a function of free-carrier density. Figure
\ref{fig:DielectricFunctionWithNe} presents the dispersion relation of Surface Plasmon Polariton and
laser wave in the case of Si, as a function of laser frequency, normalized to plasma frequency. In this calculation, the band
structure of Si has been taken into account by replacing $\varepsilon_{\infty} \left(\omega \right)$
in Eq. (\ref{eq:fulldielectricfunction}) following the measurements of Palik \cite{Palik1998}. The
wave-vector $k$ is squared and normalized to the laser pulsation to obtain the dielectric function,
and is presented as a function of pulsation $\omega$. The dispersion curve of the SPP reveals a
plasmon resonance of the free-carrier plasma if the plasma frequency is comparable with the laser
frequency e.g. if the free-carrier density is high. In the case of a doped semiconductor near the eigenfrequency, the intersection of the dispersion curves is possible \cite{Raether1986}.
In the case of femtosecond laser interaction, similarly, laser-induced ionization provides free-carrier
contribution to the dielectric function, which is equivalent to a transient doping. Therefore, the intersection also becomes possible.
The bounding of the coupled wave to the surface is described by the
condition $\omega<\frac{\omega_p}{\sqrt{2}}$ in a perfect conductor \cite{Makin2009}. In the case of damped
material, however, the leaky part of the dispersion relation between
$\omega_{spp}=\frac{\omega_{p}}{\sqrt{2}}$ and $\omega_{p}$ is allowed \cite{Maier2007}. The
condition on pulsation in a damped material is thus described by \begin{equation} \omega <
\omega_{p}. \label{eq:ConditionPulsation}\end{equation} The minimal density leading to satisfy this latter
condition is the critical density at which $\Re e \left( \varepsilon \right) = 0$.

The coupling of far field laser wave with the SPP is also possible in the case of a surface with defects
or roughness \cite{Raether1986}. In this case, the pseudo-grating period has a large
thickness $\delta k$ around its average value, and leads to a small but non-zero coupling
efficiency. Because of the laser irradiation, a coupling of a few percent is sufficient to
obtain a periodic modulation of the deposited energy by interference between laser and SPP waves
\cite{Huang2009a}.

Finally, the conditions to satisfy (Eq. \ref{eq:SPPflatCondition1}, Eq. \ref{eq:ConditionMagField},
and Eq. \ref{eq:ConditionPulsation}) allowing the excitation of the SPP at the vacuum-Si interface
can be combined into the condition: \begin{equation} \Re
e\left(\varepsilon\right)<-1\label{eq:ConditionSPP} \end{equation}

This criterion corresponds to a minimal free-carrier density given by $N_{e}=4.61 \times 10^{21}$
$cm^{-3}$ in the case of a laser irradiation of Si using a wavelength of $\lambda=800$ nm. Then, we
determine laser parameters for which this condition is justified.


\section{\label{sec:Results}Results and discussion}

\subsection{Conditions for SPP excitation on Si}

We now demonstrate that excitation of Surface Plasmon Polariton is possible on Si under femtosecond
laser irradiation, and calculate the laser parameters leading to satisfy the surface plasmon polariton
 conditions.

We now demonstrate that excitation of Surface Plasmon Polariton is possible on Si under femtosecond
laser irradiation, and calculate the laser parameters leading to satisfy the surface plasmon polariton
 conditions.
\begin{figure}[t]
	\begin{centering}
\includegraphics[width=8cm]{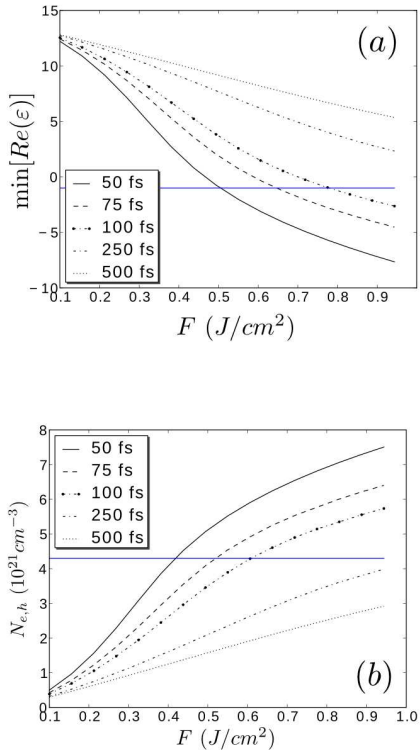}
\par\end{centering}





\begin{centering}

\par\end{centering}

\begin{centering}

\par\end{centering}

\caption{\label{fig:ResultsIndex100fs}(a) Minimum of the real part of the dielectric function as a
function of laser fluence. Blue line indicates the threshold for SPP excitation e.g. $\Re e \left( \varepsilon
\right) = -1$. (b) Maximum of the reached density during the interaction as a function of laser fluence.
Blue line indicates critical density e.g. $\Re e \left( \varepsilon \right) = 0$.  $n_{cr}=4.29 \times 10^{21} \, cm^{-3}$. Several laser pulse durations are
represented. } \end{figure} 

Figure \ref{fig:ResultsIndex100fs} (a) shows the maximum of the free-carrier density as a function
of laser fluence. At short pulse duration, the maximum free-carrier density is non-linear as a
function of fluence. These results are explained by the fact that free carrier absorption of Si
strongly depends on the carrier concentration with a resonance at the so-called critical density
\cite{Pankove1971}. The corresponding density in a solid is defined by the non-propagation condition
$\Re e \left ( \varepsilon_{Si} \right) = 0$, which gives $n_{cr}=\frac{m_e \varepsilon_0
\varepsilon_{\infty} \left( \omega^2 + \nu^2 \right) }{e^2}$. If the collision frequency $\nu=1.5 \times 10^{14} \, s^{-1}$ is
small compared to laser frequency at 800 nm wavelength $\omega=2.35 \times 10^{15} \: s^{-1}$, then, the critical density
can be expressed as $n_{cr}=\frac{m_e \varepsilon_0 \varepsilon_{\infty} \omega^2 }{e^2}$.
\cite{Pankove1971, Hulin1984, Sokolowski-Tinten2000} At low laser intensity, the maximum density
increases linearly, since the
absorption is linear in this regime. Below critical density, one observes that the efficiency of the
absorption increases, which is due to the significant contribution of the multi-photonic excitation.
Above the critical density, absorption becomes limited by the surface reflectivity. Above this
limit, the number of free-carriers still increases due to high temperature of the free-carriers,
leading to an interplay between diffusive transport and impact ionization.

Figure \ref{fig:ResultsIndex100fs} (b) shows that the resonance condition
is not met during the irradiation with a long pulse duration at the
considered fluences, since the quantity of free-carriers is limited
by the low intensity. It is shown that the real dielectric function
decreases linearly if laser fluence is near the modification
threshold ($~ 0.2$ $J/cm^{2}$, see Refs \cite{Bonse2002, Bonse2004, Korfiatis2007}). We observe that the critical density
is not reached in this fluence regime. The difference with Ref \cite{Sokolowski-Tinten2000} is explained by the different collision time, the two-photon cross section which is 10 times lower here (see Ref \cite{Bristow2007}), and the impact ionization that we took into account.  In the case of 100 fs pulse
duration, the condition given by Eq. (\ref{eq:ConditionSPP}) is satisfied
above laser fluence of 0.7 $J/cm^{2}$. A threshold for the excitation
of SPP is then identified for a laser fluence of $0.7\ J/cm^{2}$,
a pulse duration of $\tau=100$ fs, and a laser wavelength of $\lambda=800$
nm. It is also shown that under shorter laser interaction, the condition
for surface plasmon polariton excited is satisfied from lower fluences e.g.
at $0.5\ J/cm^{2}$ if $\tau=50$ fs and $\lambda=800$ nm.

From those results, a threshold fluence for SPP resonance can be defined
for each pulse duration, above which the SPP resonance conditions
are met. This result explains why a high fluence is necessary to induce
the formation of periodic structures in single pulse experiments \cite{Bonse2010,Guillermin2007,Bonse2002},
since it results in a sufficient quantity of free-carriers to excite
surface waves at the surface of Silicon.

\begin{figure}[t]
\begin{centering}
\includegraphics[width=8cm]{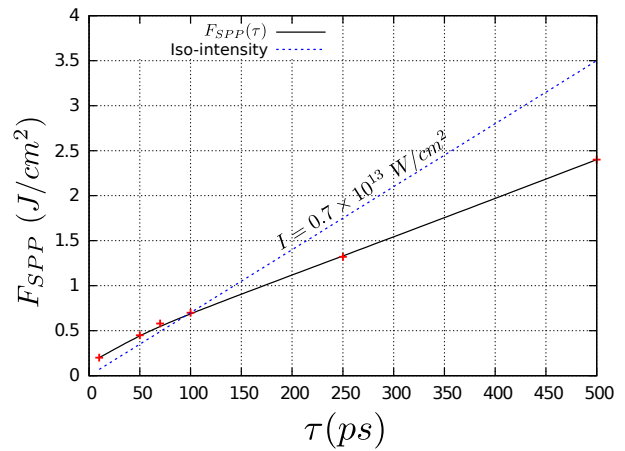}
\par\end{centering}

\caption{\label{fig:FluenceThresholdForSPP}Fluence threshold for SPP resonance
as a function of pulse duration. }
\end{figure}

In Figure \ref{fig:FluenceThresholdForSPP}, we demonstrate the fluence threshold
for the SPP resonance as a function of laser pulse duration. The corresponding
intensity, at which resonance occurs for 100 fs pulse,
is shown by the dashed curve. By comparison of the curves, we observe
that the required intensity for the SPP resonance increases with the decay in
the laser pulse duration. This effect is due to the screening and large
density gradient at the surface resulting into a strong diffusive
transport.

 Next we calculate the lifetime and the depth of the optically
active zone, e.g. the distance under the surface where the
sufficient number of free carriers are excited.
\begin{figure}[t]
\begin{centering}
\includegraphics[width=9cm]{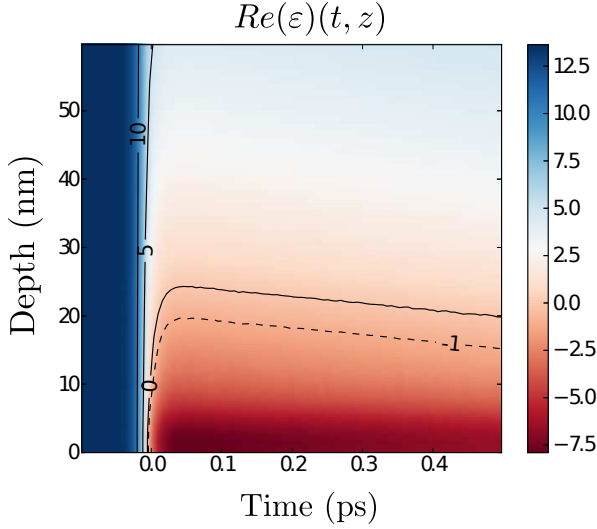}
\par\end{centering}

\caption{\label{fig:ReEpsilonDistrib50fs}Distribution of the dielectric function
as a function of depth and time. $\tau=50\ fs$, $F=0.62\ J/cm^{2}$. }
\end{figure}
Figure \ref{fig:ReEpsilonDistrib50fs} shows  the real part of the
dielectric function as a function of time and depth. It is shown that
under 50 fs pulse duration, at a fluence of $0.62\ J/cm^{2}$ , the
excited zone is nearly 20 nm deep and the SPP excitation is allowed during a
picosecond, which is greater than the pulse duration, thus leading
to the excitation of SPPs in a shorter timescale than necessary for surface melting. Moreover, the damping
length of the SPP is given by the relation \cite{Raether1986,Maier2007} $L_{SPP}=\left[2\Im m\left(\beta\right)\right]^{-1}$. The value of the damping length is contained
between $500$ nm and $2\ \text{\textmu m}$ if the SPP resonance
conditions are met. Then, the excited SPPs propagate through several micrometers
and can lead to periodic modulation if laser interferes with the
SPPs, as experimentally observed around defects \cite{Guillermin2007,Bonse2009, Bonse2010, Derrien2011, Derrien2012}. 

As underlined in the previous section, the phase-matching is
possible at the surface of Si by using a scattering
configuration with a defect or a roughness. We separate the following cases
(i) the case of the roughness, for which the size of the
scattering center is very small compared to the laser
wavelength, and (ii) the case of a defect, for which the size is
comparable to the laser wavelength. Both situations can lead to
the excitation of Surface Plasmon Polariton if the conditions
on the dielectric function are justified.

\begin{figure}[t]
\begin{centering} 
\includegraphics[height=14cm]{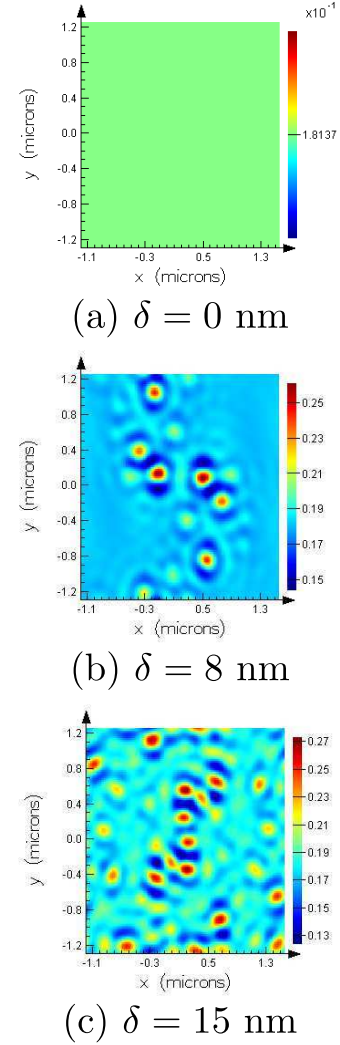}
\par\end{centering} 

\caption{\label{fig:RoughnessInfluence}Transmitted field
amplitude below the selvedge region of Si. Roughness amplitude is (a) $\delta=0$ nm, (b) $\delta=8$
nm, (c) $\delta=15$ nm at wavelength 800 nm. In these
simulations, $\varepsilon=\varepsilon_{\infty}=13.64+0.048i$. }

\end{figure} Figure \ref{fig:RoughnessInfluence} shows the
distribution of the time-averaged transmitted field amplitude and normalized by
the incident laser field intensity at the bottom of the selvedge
region. This result has been calculated using the FDTD
simulation package Lumerical \cite{Lumerical} by irradiating a
randomly generated rough Si surface using several control
parameters: FWHM of the amplitude, and distance between the
scattering centers. In this calculation, the roughness amplitude is distributed as a gaussian function between 0 and 15 nm. The distance between scattering centers is taken equal to 100 nm so that the scattered waves interfere together.
The Si dielectric constant is taken equal to
the value under 800 nm wavelength laser irradiation, in the case
of low laser excitation. One observes that the amplitude of
the field transmitted below the roughness is modulated if roughness amplitude is
greater than 8 nm. Such a roughness is formed after a single
laser pulse \cite{Torres2011} at $0.5$ $J/cm^{2}$. Thus, the amplitude allowing the
coupling of surface waves with laser is 8 nm, which explains why strictly parallel ripples are observed
after two pulses or more. Conversely, the formation of single pulse periodic structures is due to scattering
on a near-wavelength defect, which leads to the excitation of Localized
Surface Plasmon Polaritons distributed around scattering centers as
observed by several authors
\cite{Bonse2009,Guillermin2007,Derrien2011}. High fluence single
pulse experiments leads to the formation of concentric structures rather oriented
in the direction perpendicular to the laser polarization.

In this section, we have theoretically demonstrated that the excitation of Surface Plasmon
Polaritons occurs on Si irradiated by femtosecond lasers. The excitation
conditions are satisfied during the laser pulse if the laser intensity
is high. At 50 fs pulse duration, a layer of about 20 nm becomes optically
active and has a lifetime longer than the the pulse duration. We turn
now to the study of the period of the SPPs excited during  ultrashort
laser pulse on Si.

\subsection{Effect of the experimental parameters on SPP period}

The SPP period dependency on laser intensity depends on the
free-carrier density as follows
\[
\Lambda=\frac{\lambda}{\sqrt{\frac{\varepsilon_{1}\varepsilon_{2}(\omega)}{\varepsilon_{1}+\varepsilon_{2}(\omega)}}}
\]
where $\varepsilon_{1}$ and $\varepsilon_{2}(\omega)$ are respectively the dielectric functions of the media at both sides
of the vacuum - Si interface. By substituting $\varepsilon_2 (\omega)$ with Eq. (\ref{eq:fulldielectricfunction}), the periodicity of the SPP as a function of free-carrier density is calculated.

\begin{figure}
\begin{centering}
\includegraphics[width=7cm]{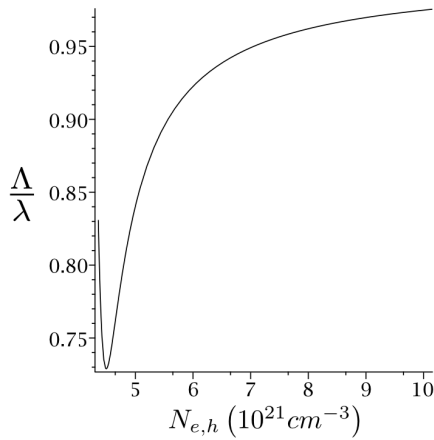}
\par\end{centering}

\caption{\label{fig:Wavelength-normalized-period}Wavelength normalized period
of SPP as a function of free-carrier density at vacuum - Si interface
when the conditions of resonance are met. }
\end{figure}
Figure \ref{fig:Wavelength-normalized-period} demonstrates the period
of the SPPs at the vacuum - Si interface as a function
of the free-carrier density. The values are presented for the free carrier number densities
required for the SPP excitation. The resulting period varies considerably with the carrier density.
The period of the SPPs is contained between $0.7\lambda$ and $\lambda$,
which correlates with the generally observed LSFL periodicities \cite{Bonse2002, Bonse2009, Bonse2010}. A
quantitative study of the variation of the SPP periodicity with laser fluence
is now presented, and compared to the LSFL ripples formed using a very
low number of laser pulses.

\begin{figure}[t]
\begin{centering}
\includegraphics[width=8cm]{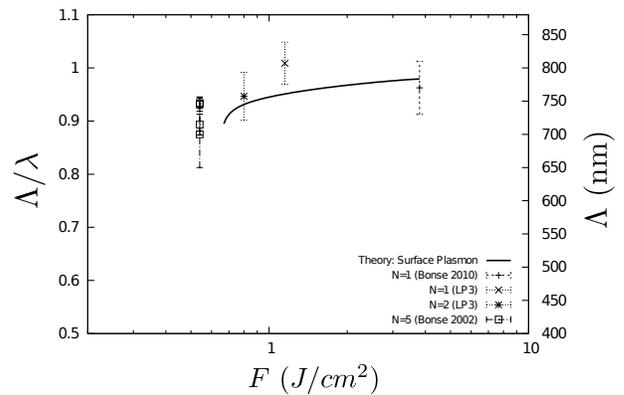}
\par\end{centering}

\caption{\label{fig:RippleThVSexp}Experimental measurements of the LSFL periods,
as a function of laser fluence. $\tau=100$ fs, $\lambda=800$ nm,
$\theta=0\text{\textdegree}$. The periods resulting from theoretical
investigations are also represented. Bonse (2010) refers to Ref. \cite{Bonse2010} and Bonse (2002) to Ref. \cite{Bonse2002}}
\end{figure}
Figure \ref{fig:RippleThVSexp} shows both theoretical and experimental periodicities. The period of the LSFL structures is presented as a function of laser fluence, for 100 fs pulse duration. When SPP
resonance conditions are satisfied, the resulting SPP period tends
to the laser wavelength when increasing laser fluence. In the
optically active range (fluence is greater than $0.7$ $J/cm^{2}$ and pulse duration $\tau=100$ fs), the calculation
results agree with the presented experimental measurements taken from Refs \cite{Bonse2002, Torres2011, Bonse2010} at very low number of pulses. The single pulse case is explained by excitation of SPP via coupling with a surface defect. The case $N=2$ is explained by coupling with roughness.
This result shows that the periodicity of Surface Plasmon Polaritons well describes the evolution
of the structure period as a function of laser fluence at reduced pulse number.

\begin{figure}[t]
\begin{centering}
\includegraphics[width=8cm]{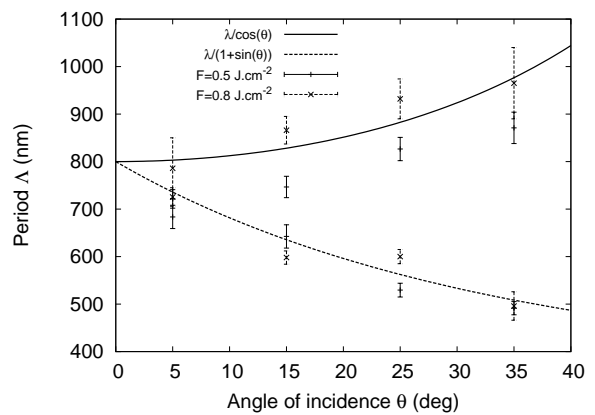}
\par\end{centering}

\caption{\label{fig:RipplePeriodAngleIncidence}Ripple periodicity as a function
of angle of incidence and laser polarization after 10 laser pulses. Pulse duration is 100 fs, and laser wavelength is 800 nm. }
\end{figure}
In the case of 10 laser pulses, LSFL ripples are well-developed, and scattering by a grating model well explains the observed periodicities, which confirms Ref \cite{Ionin2012}. Figure \ref{fig:RipplePeriodAngleIncidence} shows the comparison
between theoretical variation of period with angle of incidence with
experiments made using 10 pulses at various fluences \cite{Torres2011, Derrien2012}. Both directions
of polarization are presented. The theory of scattering by a periodically
structured surface leads to a periodicity given by \cite{Bonch-Bruevich1992}
\[
\Lambda_{P}=\frac{\lambda}{\sqrt{\eta^{2}-\sin^{2}\theta}}
\]
 where $\eta=\sqrt{\frac{\varepsilon_{1}\varepsilon_{2}}{\varepsilon_{1}+\varepsilon_{2}}}$.
 $\eta\sim1$, $\Lambda_{P}\sim\frac{\lambda}{\cos\theta}$ for
polarization parallel to the plane of incidence. For S polarization,
\[
\Lambda_{S}=\frac{\lambda}{\eta+\sin\theta}
\]
The variations of both measured and calculated ripple periods
are explained by the variation of $\eta$ near the critical
density. Actually, $\eta\sim1$ if
$Re\left(\varepsilon\right)\ll-1$, and $\eta$ varies between 0.5
and 1.5 if the condition (Eq. \ref{eq:ConditionSPP}) is
satisfied. This section demonstrated that the theory of
SPP excitation on gratings agrees with the experiments and
explains the variation of the ripple period with the angle of
incidence and with laser polarization. Such a modulation arises after several pulses. Several initial laser pulses generate roughness. The next pulses excite SPPs by coupling with surface roughness and defects, then the interference of SPP with laser pulse leads to periodic modulation of the energy in the time scale of the laser pulse. Thus, a periodic phase transition is achieved by electron-phonon coupling, and leads to the structuring of the surface, following a pattern given by SPP periodicity. \cite{Derrien2011, Colombier2012}. We underline that the
threshold fluence allowing SPP resonance on the surface of a grating is
decreased with respect to the results presented in Figure
\ref{fig:ResultsIndex100fs} (a), since the energy absorption is enhanced in the presence of the grating \cite{Ursu1984}.

\section{Conclusions}

The possibility of the surface plasmon polariton excitation on Si
surface irradiated by a femtosecond laser pulse has been theoretically demonstrated.
A sufficient number of free-carriers is excited from the valence band
to the conduction band during the laser pulse, thus satisfying the SPP excitation conditions. The required
ranges of laser fluences and pulse durations have been identified
to satisfy the SPP excitation conditions. In particular,
SPPs can be excited by using a femtosecond laser with 800 nm wavelength, 100 fs pulse duration and with laser fluences larger
than $0.7$ $J/cm^{2}$. As a result, a thin layer is excited, with a lifetime longer than the pulse duration and with
a depth of several tens of nanometers. The threshold intensity required to excite SPPs is higher with a sub-100 fs pulse duration than with a pulse duration longer than 100 fs. This effect is due to the free-carrier diffusion induced by strong gradients of free-carrier density and energy,  enhanced at low pulse duration, and to the increase of surface reflectivity which limits the absorbed energy at short pulse duration. 

Furthermore, a comparison of the calculated SPP periodicities and experimentally measured ripple
periodicities allows us to conclude that the formation
of periodic structures with a reduced number of laser pulses is due to the excitation of SPPs at the Si surface.

The presence of a surface roughness with $\delta\ll\lambda$ leads
to the coupling of the laser wave with the roughness. We have found
that the required roughness amplitude allowing the coupling of laser
wave with the surface is 8 nm. It is also possible to obtain periodic
structures by scattering on defects ($\delta\sim\lambda$) that are
present at the surface by excitation of localized surface plasmon polaritons.
These results underline the importance of the surface quality in the
SPP excitation and thus to the LSFL ripple formation.

As a result of the performed analysis, the possibilities of control over the
period of the LSFL ripples can be deduced in the regime of low number of pulses.
The period can be reduced down to 40\%  by increasing
the angle of incidence for S polarization from normal incidence to $40 \text{\textdegree}$ incidence, and can be increased up
to 37\% by increasing the angle of incidence for P polarization from normal incidence to $40 \text{\textdegree}$ of incidence. The
period of the LSFL ripples can be increased of 10\% by increasing the laser fluence from excitation
threshold $0.7\: J/cm^{2}$ up to $5\: J/cm^{2}$. Finally, the period of the LSFL structure is shown
to be reduced with the number of laser pulses and can be decreased by $50$ \% with respect to the
one obtained for a single pulse. However, the corresponding mechanism is still under discussions
\cite{Bonse2010, Tsibidis2011}. Our calculation results have thus demonstrated the existence of the
lower fluence limit, below which the surface plasmon polaritons can not be excited on Si. In
addition, the model has an upper fluence limit above which the band structure is destroyed and the
material is severely damaged or ablated. Taking into account that for ripple formation, laser wave
should enter in resonance with SPP wave, we confirm the fact that there is a well-defined fluence
window for ripple formation due to SPP \cite{Bonse2009}, which depends on laser wavelength and pulse
duration. For the considered parameters, fluence is in the range between $0.7$ $J/cm^{2}$ and $5$
$J/cm^{2}$.

\section*{Acknowledgments}

TJD is grateful to the French Ministry of Research for the PhD grant. The National Computational Center for Higher Education (CINES) under project c2011085015 is acknowledged.


\begin{thebibliography}{10}

\bibitem{Birnbaum1965}
M.~Birnbaum.
\newblock Journal of Applied Physics \textbf{36}, 11, p. 3688 (1965).

\bibitem{Young1983}
J.~F. Young, J.~Preston, H.~V. Driel, and J.~Sipe.
\newblock Physical Review B \textbf{27}, p.~2 (1983).

\bibitem{Ursu1985}
I.~Ursu, I.~Mihailescu, L.~Nistor, V.~Teodorescu, A.~Prokhorov, V.~Konov, and
  V.~Tokarev.
\newblock Applied Optics \textbf{24}, p.~22 (1985).

\bibitem{Bonse2010}
J.~Bonse and J.~Kr\"uger.
\newblock Journal of Applied Physics \textbf{108}, p. 034903 (2010).

\bibitem{Tsibidis2012}
G.~Tsibidis, E.~Stratakis, and K.~Aifantis.
\newblock Journal of Applied Physics \textbf{111}, p. 053502 (2012).

\bibitem{Renger2009}
J.~Renger, R.~Quidant, N.~van Hulst, S.~Palomba, and L.~Novotny.
\newblock Physical Review Letters \textbf{103}, p. 266802 (2009).

\bibitem{Borowiec2003}
A.~Borowiec.
\newblock Applied Physics Letters \textbf{82}, p.~25 (2003).

\bibitem{Bonse2005}
J.~Bonse, M.~Munz, and H.~Sturm.
\newblock Journal of Applied Physics \textbf{97}, p. 013538 (2005).

\bibitem{Vorobyev2008a}
A.~Vorobyev and C.~Guo.
\newblock Journal Of Applied Physics \textbf{104}, p. 063523 (2008).

\bibitem{Crawford2007}
T.~Crawford and H.~Haugen.
\newblock Applied Surface Science \textbf{253}, pp. 4970--4977 (2007).

\bibitem{Vorobyev2008}
A.~Vorobyev.
\newblock Journal of Applied Physics \textbf{104}, p. 053516 (2008).

\bibitem{Vorobyev2008b}
A.~Vorobyev and C.~Guo.
\newblock Applied Physics Letters \textbf{92}, p. 041914 (2008).

\bibitem{Reif2002}
J.~Reif, F.~Costache, M.~Henyk, and S.~V. Pandelov.
\newblock Applied Surface Science \textbf{197-198}, pp. 891--895 (2002).

\bibitem{Ben-Yakar2007}
A.~Ben-Yakar, A.~Harkin, J.~Ashmore, R.~Byer, and H.~Stone.
\newblock Journal of Physics D: Applied Physics \textbf{40}, p. 1447 (2007).

\bibitem{Varlamova2007}
O.~Varlamova, F.~Costache, M.~Ratzke, and J.~Reif.
\newblock Applied Surface Science \textbf{253}, pp. 7932--7936 (2007).

\bibitem{Her1998}
T.~Her, R.~Finlay, C.~Wu, S.~Deliwala, and E.~Mazur.
\newblock Applied Physics Letters \textbf{73}, p.~12 (1998).

\bibitem{Carey2003}
J.~E. Carey, C.~H. Crouch, and E.~Mazur.
\newblock Optics and Photonics News \textbf{14}, p.~2 (2003).

\bibitem{Sarnet2008a}
T.~Sarnet, R.~Torres, V.~Vervisch, P.~Delaporte, M.~Sentis, M.~Halbwax,
  J.~Ferreira, D.~Barakel, M.~Pasquinelli, S.~Martinuzzi, L.~Escoubas,
  F.~Torregrosa, H.~Etienne, and L.~Roux.
\newblock ICALEO 2008 Congress Proceedings \textbf{101}, p. 161 (2008).

\bibitem{Reif2010}
J.~Reif, O.~Varlamova, M.~Ratzke, M.~Schade, H.~Leipner, and T.~Arguirov.
\newblock Applied Physics A \textbf{101}, p. 361 (2010).

\bibitem{Sipe1983}
J.~Sipe, J.~F. Young, J.~Preston, and H.~V. Driel.
\newblock Physical Review B \textbf{27}, p.~2 (1983).

\bibitem{Skolski2012}
J.~Skolski, G.~R\"omer, J.~Obona, V.~Ocelik, A.~H. in't Veld, and J.~T. M.~D.
  Hosson.
\newblock Physical Review B \textbf{85}, p. 075320 (2012).

\bibitem{Bonse2009}
J.~Bonse, A.~Rosenfeld, and J.~Kr\"uger.
\newblock Journal of Applied Physics \textbf{106}, p. 104910 (2009).

\bibitem{Costache2004}
F.~Costache.
\newblock Applied Physics A \textbf{79}, p. 1429 (2004).

\bibitem{Reif2006a}
J.~Reif, M.~Ratzke, O.~Varlamova, and F.~Costache.
\newblock Materials Science and Engineering B \textbf{134}, pp. 114--117
  (2006).

\bibitem{Tsibidis2011}
G.~Tsibidis, M.~Barberoglou, P.~Loukakos, E.~Stratakis, and C.~Fotakis.
\newblock Physical Review B \textbf{86}, p. 115316 (2012).

\bibitem{Guillermin2007}
M.~Guillermin, F.~Garrelie, N.~Sanner, E.~Audouard, and H.~Soder.
\newblock Applied Surface Science \textbf{253}, pp. 8075--8079 (2007).

\bibitem{Miyaji2008}
G.~Miyaji and K.~Miyazaki.
\newblock Optics Express \textbf{16}, p.~20 (2008).

\bibitem{Huang2009a}
M.~Huang, F.~Zhao, Y.~Cheng, N.~Xu, and Z.~Xu.
\newblock ACS Nano \textbf{3}, p.~12 (2009).

\bibitem{Sakabe2009}
S.~Sakabe.
\newblock Physical Review B \textbf{79}, p. 033409 (2009).

\bibitem{Garrelie2011}
F.~Garrelie, J.~Colombier, F.~Pigeon, S.~Tonchev, N.~Faure, M.~Bounhalli,
  S.~Reynaud, and O.~Parriaux.
\newblock Optics Express \textbf{19}, p.~10 (2011).

\bibitem{Raether1986}
H.~Raether.
\newblock \emph{Surface plasmons on smooth and rough surfaces and on gratings}
  (Springer-Verlag, 1986).

\bibitem{Zayats2005}
A.~Zayats, I.~Smolyaninoy, and A.~Maradudin.
\newblock Physics Reports \textbf{408}, pp. 131--314 (2005).

\bibitem{Bonse2011}
J.~Bonse, A.~Rosenfeld, and J.~Kr\"uger.
\newblock Applied Surface Science \textbf{257}, pp. 5420--5423 (2011).

\bibitem{Hecht1996}
B.~Hecht, H.~Bielefeldt, L.~Novotny, Y.~Inouye, and D.~Pohl.
\newblock Physical Review Letters \textbf{77}, p.~9 (1996).

\bibitem{Sokolowski-Tinten1998}
K.~Sokolowski-Tinten, J.~Bialkowski, A.~Cavalleri, D.~V.~D. Linde, A.~Oparin,
  J.~M. ter Vehn, and S.~Anisimov.
\newblock Physical Review Letters \textbf{81}, p.~1 (1998).

\bibitem{Sokolowski-Tinten2000}
K.~Sokolowski-Tinten and D.~von~der Linde.
\newblock Physical Review B \textbf{61}, p.~4 (2000).

\bibitem{Bok1981}
J.~Bok and M.~Combescot.
\newblock Physical Review Letters \textbf{47}, p.~21 (1981).

\bibitem{Driel1987}
H.~V. Driel.
\newblock Physical Review B \textbf{35}, p.~15 (1987).

\bibitem{Sjodin1998}
T.~Sjodin, H.~Petek, and H.-L. Dai.
\newblock Physical Review Letters \textbf{81}, p.~25 (1998).

\bibitem{Palik1998}
E.~Palik.
\newblock \emph{Handbook of Optical Constants of Solids} (Academic Press,
  1985).

\bibitem{Sze2007}
S.~Sze and K.~K. Ng.
\newblock \emph{Physics Of Semiconductor Devices} (Wiley-Interscience, 2007).

\bibitem{Thornber1981}
K.~Thornber.
\newblock Journal Of Applied Physics \textbf{52}, p. 279 (1981).

\bibitem{Choi2002a}
T.~Y. Choi and C.~P. Grigoropoulos.
\newblock Journal of Applied Physics \textbf{92}, p.~9 (2002).

\bibitem{Bulgakova2005}
N.~Bulgakova, R.~Stoian, A.~Rosenfeld, I.~Hertel, W.~Marine, and E.~Campbell.
\newblock Applied Physics A \textbf{81}, pp. 345--356 (2005).

\bibitem{Bonse2004}
J.~Bonse, K.~Brzezinka, and A.~Meixner.
\newblock Applied Surface Science \textbf{221}, pp. 215--230 (2004).

\bibitem{Fischetti1988}
M.~Fischetti and S.~Laux.
\newblock Physical Review B \textbf{38}, p.~14 (1988).

\bibitem{Born1980}
M.~Born and E.~Wolf.
\newblock \emph{Principles of Optics. Electromagnetic theory of propagation,
  interference and diffraction of light.} (Cambridge University Press, 1980),
  7th edition edition.

\bibitem{Baeuerle2000}
D.~B\"auerle.
\newblock \emph{Laser Processing and Chemistry} (Springer, 2000).

\bibitem{Bulgakova2010}
N.~Bulgakova, R.~Stoian, and A.~Rosenfeld.
\newblock Quantum Electronics \textbf{40}, p.~11 (2010).

\bibitem{Jackson1999}
J.~Jackson.
\newblock \emph{Classical electrodynamics} (Wiley, 1999).

\bibitem{Jou1993}
D.~Jou, J.~Casas-Vazquez, and G.~Lebon.
\newblock \emph{Extended irreversible thermo-dynamics} (Springer-Verlag, 1993).

\bibitem{Alvarez2007}
F.~Alvarez and D.~Jou.
\newblock Applied Physics Letters \textbf{90}, p. 083109 (2007).

\bibitem{Palpant2008}
B.~Palpant, Y.~Guillet, M.~Rashidi-Huyeh, and D.~Prot.
\newblock Gold Bulletin 2008 \textbf{41}, p.~2 (2008).

\bibitem{Desai1986}
P.~Desai.
\newblock Journal of Chemical Reference and Datas \textbf{15}, p.~3 (1986).

\bibitem{Rhim2000}
W.-K. Rhim and K.~Ohsaka.
\newblock Journal of Crystal Growth \textbf{208}, pp. 313--321 (2000).

\bibitem{Magna2004}
A.~L. Magna, P.~Alippi, V.~Privitera, G.~Fortunato, M.~Camalleri, and
  B.~Svensson.
\newblock Journal Of Applied Physics \textbf{95}, p.~9 (2004).

\bibitem{Rousse2001}
A.~Rousse, C.~Rischel, S.~Fournaux, I.~Uschmann, S.~Sebban, G.~Grillon,
  P.~Balcou, E.~Forster, J.~Geindre, P.~Audebert, J.~Gauthier, and D.~Hulin.
\newblock Nature \textbf{410}, pp. 65--68 (2001).

\bibitem{Sundaram2002}
S.~Sundaram and E.~Mazur.
\newblock Nature Materials \textbf{1}, p. 217 (2002).

\bibitem{Sokolowski-Tinten2003}
K.~Sokolowski-Tinten, C.~Blome, J.~Blums, A.~Cavalleri, C.~Dietrich,
  A.~Tarasevitch, I.~Uschmann, E.~Forster, M.~Kammler, M.~H. von Hoegen, and
  D.~von~der Linde.
\newblock Nature \textbf{422}, pp. 287--289 (2003).

\bibitem{Combescot1985}
M.~Combescot and J.~Bok.
\newblock Journal of Luminescence \textbf{30}, pp. 1--17 (1985).

\bibitem{Ursu1984}
I.~Ursu, I.~Mihailescu, A.~Popa, A.~Prokhorov, V.~Konov, V.~Ageev, and
  V.~Tokarev.
\newblock Applied Physics Letters \textbf{45}, pp. 365--367 (1984).

\bibitem{Bonch-Bruevich1992}
A.~M. Bonch-Bruevich, M.~N. Libenson, V.~S. Makin, and V.~A. Trubaev.
\newblock Optical Engineering \textbf{31(4)}, pp. 718--730 (1992).

\bibitem{Novotny1997}
L.~Novotny, B.~Hechtt, and D.~Pohl.
\newblock Journal of Applied Physics \textbf{81}, p.~4 (1997).

\bibitem{Maier2007}
S.~A. Maier.
\newblock \emph{Plasmonics, Fundamentals and Applications} (Springer, 2007).

\bibitem{Makin2009}
V.~Makin, Y.~I. Pestov, R.~Makin, and A.~Y. Vorobyev.
\newblock Journal of Optical Technology \textbf{76}, p.~9 (2009).

\bibitem{Pankove1971}
J.~I. Pankove.
\newblock \emph{Optical Processes in Semiconductors} (Dover Science, 1971).

\bibitem{Hulin1984}
D.~Hulin, M.~Combescot, J.~Bok, A.~Migus, J.~Y. Vinet, and A.~Antenotti.
\newblock Physical Review Letters \textbf{52}, p.~22 (1984).

\bibitem{Bonse2002}
J.~Bonse, S.~Baudach, J.~Kr\"uger, W.~Kautek, and M.~Lenzner.
\newblock Applied Physics A \textbf{74}, pp. 19--25 (2002).

\bibitem{Korfiatis2007}
D.~Korfiatis, K.~Thoma, and J.~Vardaxoglou.
\newblock Journal of Physics D: Applied Physics \textbf{40}, p. 6803 (2007).

\bibitem{Bristow2007}
A.~D. Bristow, N.~Rotenberg, and H.~M.~V. Driel.
\newblock Applied Physics Letters \textbf{90}, p. 191104 (2007).

\bibitem{Derrien2011}
T.~J.-Y. Derrien, R.~Torres, T.~Sarnet, M.~Sentis, and T.~Itina.
\newblock Applied Surface Science \textbf{258}, p.~23 (2012).

\bibitem{Derrien2012}
T.~J.-Y. Derrien.
\newblock \emph{Nanostructuration de cellules photovoltaiques par impulsion
  laser femtoseconde. Etude des m\'{e}canismes de formation.}
\newblock Ph.D. thesis, Universit\'{e} de la M\'{e}diterran\'{e}e - Aix
  Marseille II (2012).

\bibitem{Lumerical}
Lumerical.
\newblock FDTD Solutions, http://www.lumerical.com.

\bibitem{Torres2011}
R.~Torres.
\newblock \emph{Structuration du silicium par laser femtoseconde : application
  au photovolta\"ique}.
\newblock Ph.D. thesis, Universit\'e de la M\'editerran\'ee (2011).

\bibitem{Ionin2012}
A.~A. Ionin, S.~I. Kudryashov, S.~Makarov, L.~Seleznev, D.~Sinitsyn,
  E.~Golosov, O.~A. Golosova, U.~Kobolov, and A.~Ligachev.
\newblock Applied Physics A \textbf{107}, pp. 301--305 (2012).

\bibitem{Colombier2012}
J.~Colombier, F.~Garrelie, N.~Faure, S.~Reynaud, M.~Bounhalli, E.~Audouard,
  R.~Stoian, and F.~Pigeon.
\newblock Journal Of Applied Physics \textbf{111}, p. 024902 (2012).

\end{thebibliography}

\listoffigures

\end{document}